# Improving Reliability of Human Trafficking Alerts in Airports

Nana Oye Akrofi Quarcoo
*School of Computer Science*
*The University of Nottingham*
psxna21@nottingham.ac.uk

Milena Radenkovic
*School of Computer Science*
*The University of Nottingham*
milena.radenkovic@nottingham.ac.uk

*Abstract*— **This paper investigates the latter scenario of individual emergency alerts in airports by applying two existing benchmark delay tolerant network protocols and evaluating their performance of delivery ratio and latency. First, the paper provides a background on Mobile Ad Hoc Networks (MANETs) and Delay Tolerant Networks (DTNs), as well as Vehicular Ad Hoc Networks (VANETs) as a subset of MANETs. Next, the scenario is simulated using the Opportunistic Network Environment (ONE) simulator and runs the DTN protocols applying Spray and Wait and Epidemic. The study discusses the results, highlighting the advantages and limitations of each protocol within the scenario and addressing constraints of the simulation or experimental setup. A wider discussion then considers related research on technologies that combat human trafficking and the potential role of DTN networks in improving this global issue for the better.**

*Keywords — DNT, MANET, Human trafficking*

## I. INTRODUCTION

Airports worldwide are drawing urgent attention to the critical global crisis of human trafficking, urging at-risk individuals to alert authorities[16]. In scenarios where conventional networks like cellular or Wi-Fi are unavailable, do we leverage direct device-to-device communication to ensure connectivity and data exchange?

This paper investigates the latter scenario of individual emergency alerts in airports by applying two existing benchmark delay tolerant network protocols and evaluating their performance of delivery ratio and latency. First, the paper provides a background on Mobile Ad Hoc Networks (MANETs) and Delay Tolerant Networks (DTNs), as well as Vehicular Ad Hoc Networks (VANETs) as a subset of MANETs. Next, the scenario is simulated using the Opportunistic Network Environment (ONE) simulator [11] and runs the DTN protocols applying Spray and Wait and Epidemic. The study discusses the results, highlighting the advantages and limitations of each protocol within the scenario and addressing constraints of the simulation or experimental setup. A wider discussion then considers related research on technologies that combat human trafficking and the potential role of DTN networks in improving this global issue for the better.

## II. MOBILE AD HOC NETWORKS (MANETs) AND VEHICLE AD HOC NETWORKS (VANETs)

### A. MANETs

In everyday life, mobile devices are playing an important role in our daily lives and billions of users rely on cellular (4g/5g) or local networks like WIFI routers to browse the internet or use various applications services like social media. GSMA [8] reports that there are approximately 5.79 billion unique mobile subscribers worldwide as of Q4 2025, representing about 71% of the global population. Most of these devices utilize fixed-infrastructure and show how network services are in highest demand. Further, this creates a challenge in the costs and times to set up the infrastructure network [5]. Another challenge occurs in situations where user-required infrastructure e.g cellular networks are not available or can be installed within a given geographic area due to e.g disasters or war. Providing the needed connectivity and network services in these situations among the challenges presented, is what mobile ad hoc networking is designed for [5]. Mobile ad hoc networks are self-configuring networks, meaning that mobile devices are connected via wireless links e.g. bluetooth without a fixed infrastructure. The nodes behave as routers, and take part in discovery and maintenance of routes to other nodes in the local network[5]. The notation of behaving as routers refers to the nodes being able to forward their own data, but also forward data to a destination in which other nodes cannot reach directly. Here, they use multi-hop communication [5], where a node A wants to send a message to a destination D, but before reaching it, it has to travel through multiple hops. In terms of mobility they are free to move randomly at any time and organize themselves arbitrarily, hence the topology is defined as dynamic in nature [5]. Applications where such networks are found are in for example military communication, disaster recovery or sensor networks. In the context of military operations, we see that soldiers or vehicles move constantly, so the network layout keeps changing. Furthermore, it is not possible to rely on fixed infrastructure for several reasons; for example, natural disasters like earthquakes, hurricanes or fire can destroy infrastructure, power may not exist in combat areas or terrain may prevent installation like sensor networks in some rural areas. Sensor networks are also an example of MANETs, which can consist of 1000 - 1000.000 sensors that collect data and forward it to a centralized host for processing and utilizing it for a specific service [5].

### B. VANETs

A specialized type of MANET's is Vehicle Ad Hoc Networks (VANET's) which are specifically designed for vehicles, that could be cars, drones etc. [7]. They have a different mobility pattern in terms of speeds and the network topology is more frequent and dynamic due to their mobility. However, movement within the network can be more predictable due to predefined paths, routines, or scheduled movements.

Research within VANET is focused on Intelligent Transportation Systems (ITS), that aims to for example reduce traffic congestion, road accidents and similar [7]. The latter require different forms of communication. For example an application concerning collision avoidance, vehicles must utilize vehicle-to-vehicle communication to communicate within milliseconds, or the warning may arrive too late to prevent an accident. For this, vehicles use wireless communication interfaces e.g 802.11p[7]. The network changes quickly, so the Quality of Service (QoS) must be very high with low delay and reliable communication for safety, unlike MANETs where some delays might be tolerated in some situations. There is also vehicle-to-roadside(V2R) communication that exploits infrastructure-based wireless technologies to communicate with roadside units. That could be information about traffic information or parking availability. MANETs and VANETs face similar challenges, including lack of infrastructure, rapidly changing network topologies, limited wireless range and bandwidth [7]. However, their characteristics lead to different ways of solving these challenges leading to different protocols of MANET's and VANETS. For example, when looking at energy constraints, MANET's are often battery powered where vehicles have a more stable power source. So in this aspect, VANET protocols may have to focus less on energy constraints.

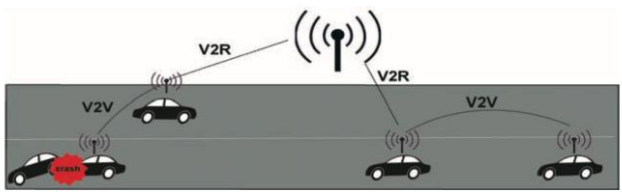

Fig. 1. Illustration of VANET communication concepts.

## III. DELAY TOLERANT NETWORKS (DTNs)

MANET's route discovery assumes that there is an existing route from source to destination; when a node wants to send a packet, it discovers the path ( on-demand routing) or looks up the path in the routing table that it maintains (proactive routing[18]. While forwarding the message, if a link between two nodes fails e.g nodes move out of range, then the package is usually dropped. Delay tolerant networks is an advanced way of optimizing MANET's in terms of this problem - what if the "next hop" doesn't exist? How do we then ensure that the message is not dropped? DTN addresses challenges in disconnected and disrupted networks that assume no end-to-end connection [18]. They are designed to operate effectively in contexts where extreme distances or latency occurs such as space communications or an interplanetary scale [18]. When a node stores a message, it carries it while moving physically or vehicularly and forwards it later when a connection becomes available. The latter is described as a store-carry-forward approach. DTN protocols classifications may be split up in Replication-based and Forwarding-based protocols[18]. While replication based spreads messages to neighbors throughout the network (flooding) it increases the probability of reaching destination D because multiple copies of the message exist in the network [18]. However some research issues challenges in replication based routing schemes such as network congestion in clustered areas and strain on network resources regarding bandwidth, storage and energy[18]. Forwarding-based protocols maintain only a single copy of a message at each node at any given time[18]. When the destination is reached upon a source sending the message, no other node can have a copy, which eliminates the need for providing feedback to the network about deleting copies [18]. Forwarding based protocols deal with the challenge of flooding by limiting the use of resources by selecting the best paths to a destination. Although they cost much less network resources, new challenges arise in terms of the time it may take to establish these paths. Overall, each routing approach has its advantages and disadvantages, and the choice of a specific protocol depends on the context and the problem scenario, as well as which trade-offs a system can tolerate. For example, in the context of human trafficking prevention in airports, a slightly delayed message from an at-risk individual still allows staff to respond, coordinate with other teams and communicate with nearby airports. Here, the priority would be to ensure that the messages are reliably delivered, even if not immediately.

## IV. DTN PROTOCOLS OVERVIEW

### A. Epidemic

In epidemic routing every node in the network will transmit messages to each other upon an encounter, if the encountering node does not possess the copy of the message message [17]. The protocol relies on carriers(nodes) within the network to come into contact with another part of the network that is connected through mobility[17].

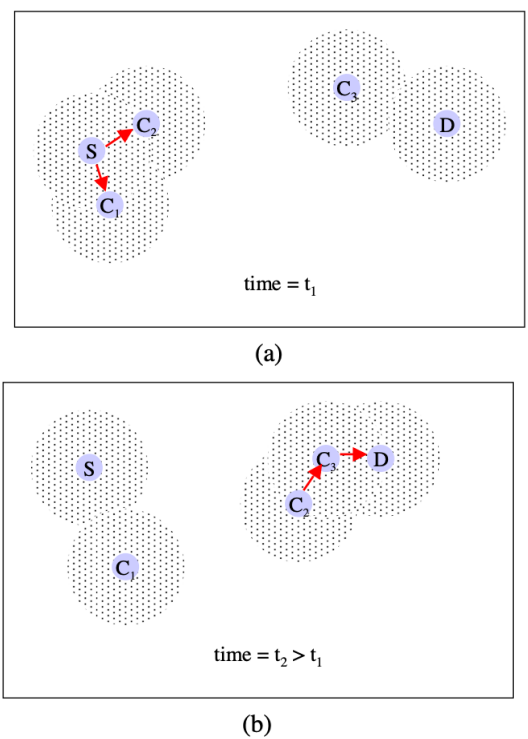

Fig. 2. S (source) wants to send a message to D (destination). Initially, S and D do not have a connected path, but later C2 moves and encounters C3, which has a direct connection to D.

As nodes repeatedly exchange messages and through mobility, the cascading spread of message copies ensures a high probability that the destination will be reached[ref]. Every node stores and forwards the message regardless of whether it has a realistic chance of helping the message reach its destination[17]. The main goal of epidemic routing is to obtain high delivery rates and minimize delivery latency[17].

*B. Spray and Focus*

Controlled replication or spraying is a routing method where the aim is to control and limit the number of messages in the network. Controlled replication aims for high delivery ratios but by reducing overhead, that means decreasing extra work, traffic, or resource usage that is not directly contributing to delivering the actual message[15].

*1) Spray and Wait*

Spray and Wait is an example of a controlled replication. It creates low resource efficiency by setting a strict limit of the number of copies per message allowed in the network[15]. The protocol consists of two phases, the spray and the wait phase. When a source sends a message, a number L is attached defining the maximum allowable copies of the message in the network. During "spray", the source delivers one copy to the L distinct nodes. The node then enters the wait phase, where the node holds the message until the destination is reached directly[15]. The latter refers to the vanilla version of the protocol.

A binary version similarly starts with L copies of the message, but transmits half of the total number of copies it has. Each of the nodes then transfers half of the messages to the next encounter and so one, until it has one. At that moment, it switches to the wait phase, and as the vanilla version, waits for an opportunity of direct transmission to the destination. The benefit in the binary version is that messages are spread faster across the network, because one node is able to give out more messages due to the amount of copies[15].

*2) Spray and Focus (SnF)*

It has shown that spraying performs in scenarios with high mobility like in VANET's as they encounter far many nodes. However, they struggle when mobility is slow; they perform poorly when nodes move slowly or tend to travel together along similar paths, limiting the opportunities for message forwarding. The spray and focus protocols aims to solve this challenge and provides a scheme where each node can forward its copy, using a forwarding based scheme, instead of naively waiting to deliver it to the destination itself [15]. In the spraying phase, when a new message is generated at a source note, L "forwarding tokens" is chosen. If the node has more than one token, it performs binary spray, that is forwarding half of the copies. If the node has exactly one forwarding token, it performs Utility Based Forwarding, with the last encounter times used as the function [15]. The timers keep track of when nodes encounter each other within communication range [15]. That means, a node can look at all neighbours within its communication range and check their last encounter timers relative to the destination. Based on that, it can make the decision of forwarding to the most optimal node.

## V. SCENARIO EXPERIMENT: HUMAN TRAFFICKING IN AIRPORTS

*A. A critical, global challenge*

Human trafficking is the fastest-growing and the second-largest criminal industry globally and in 2018 [4], an estimated 24.9 million people were illegally trafficked and live in conditions of modern slavery[4]. Human trafficking is defined as:

*"The recruitment, transportation, transfer, harbouring, or receipt of people through force, fraud, or deception, with the aim of exploiting them for profit."*

This represents a significant global challenge , one that is likely to intensify as climate change, conflicts or wars, and other crises continue to escalate which creates rich opportunities for human trafficking[10] International organizations are raising alarms about this concern in a number of other conflict-related crises, including in the Central African Republic, Colombia, Democratic Republic of the Congo, Iraq, Libya, Mali, Myanmar, Somalia, South Sudan, Sudan, Syrian Arab Republic, and Yemen[PDF]. The International Air Transport Association(IATA) further concludes that 80% of human trafficking routes pass through official border controls, including airports[4]. In response, the UNODC has developed training programs for airport staff to identify and respond to potential cases of human trafficking which is adopted by airports worldwide, showcasing its challenge and importance of attention.

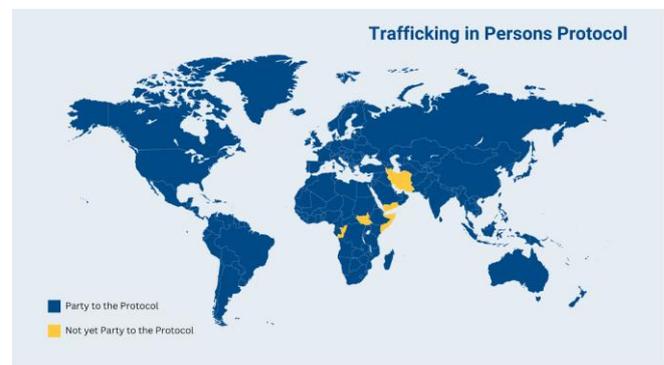

Fig. 3. The protocol: a legally binding international treaty that defines human trafficking, requires countries to criminalize it, and mandates cooperation in prevention, prosecution, and protection of victims.

In addition, awareness campaigns, such as informational posters aim to alert travellers and staff to the issue[14] 4.

Several airports have introduced similar visual campaigns, placing it in restrooms, arrival halls, baggage areas, and airside zones where potential victims may have brief moments away from traffickers. These materials often include discreet hotline numbers, multilingual messages, and clear guidance on what suspicious behaviours to look for[14].

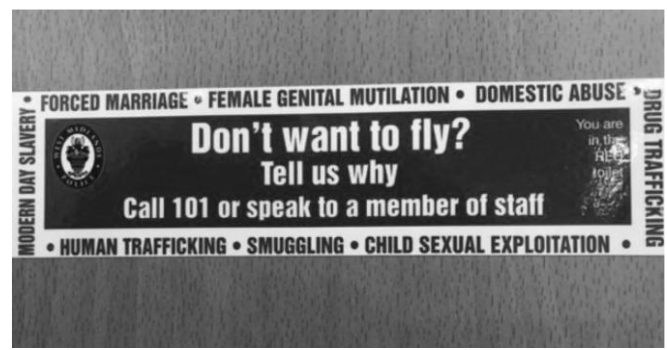

Fig. 4. Stickers displayed in toilets at Birmingham Airport ask passengers at risk of trafficking or modern-day slavery to alert for help

## B. Experiment setup and configuration

### 1) Scenario-based design

Scenario-based design allows us to explicitly describe how actors interact within a specific context, focusing on their activities and goals rather than system functions[6]. By defining node groups, movements, and interactions, we can create a concrete yet flexible representation of an airport environment. This approach enables testing and evaluation of different routing protocols in a controlled, scalable simulation. The scenario is set in a generic indoor airport, including pedestrians (travellers), security personnel, and static locations such as shops and restrooms. Due to the absence of detailed data on categories, the model focuses on the latter-described selection - passengers, staff, and security personnel—although other elements, such as trams, transport vehicles, or additional facilities, could also be incorporated.

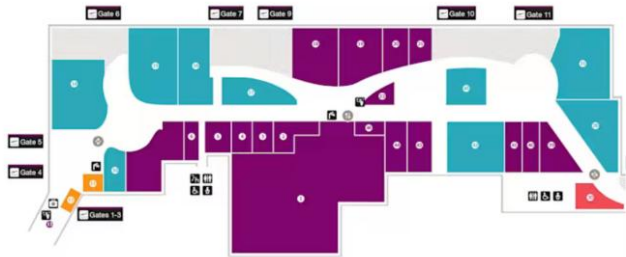

Fig. 5. Edinburgh airport

### 2) The One Simulator: Simulation Tool for Mobile Ad Hoc Networks.

The Opportunistic Network Environment simulator (The ONE) is used to provide a realistic environment for configuring and evaluating real world DTN applications and scenarios [11]. It enables the setup of scenarios with configurable parameters, including mobility patterns, event generation, message exchange, communication types, real data traces and routing protocols [11].

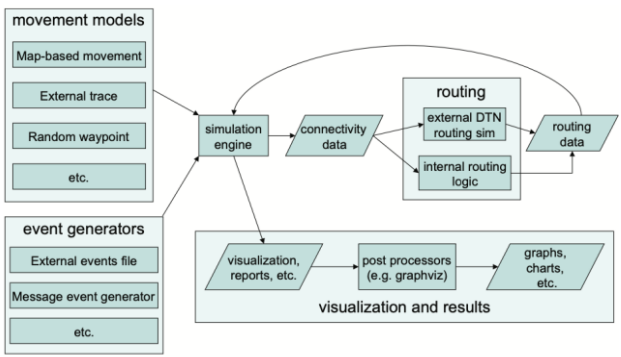

Fig. 6. Overview of the ONE simulation environment

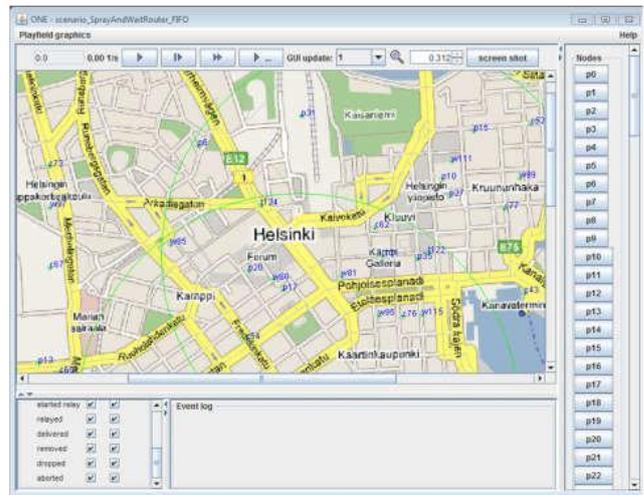

Fig. 7. The ONE graphical user interface (GUI)

Additionally, the simulator generates a report containing the results of each simulation run, which is used for analysis. The simulator provides a graphical user interface(GUI) that shows the stimulation state; the locations and the node movement with messages carried by the nodes, as well as active contacts[11]We will use the One simulator to set up the problem scenario described in the motivation and apply and evaluate Epidemic and Spray and Focus on this problem.

### 3) Node group configuration

This section describes the configuration of the node groups used in the simulation.

In the absence of data traces from an airport, the spatial layout of nodes follows the Helsinki downtown map, which is default in the ONE simulator as seen in figure 7. This includes roads and pedestrian walkways, reflecting typical pedestrian movement patterns. Using this map as a reference, gates, shops, and restrooms are positioned to be spread out across the area, mimicking the general distribution of facilities in an airport, as illustrated in 5It is important to note that this setup does not model any specific real world airport; the layout is only inspired by typical airport features. Each simulation runs for 14,400 seconds (4 hours), providing time to observe how messages propagate through the network. Messages are generated repeatedly by the event generator at intervals of 25–35 seconds. Two designated source nodes transmit HELP messages to destination hosts (all security nodes).

| Group | Role | # of Hosts | Movement Model | Speed | Wait Time | Buffer Size | Interface | Cluster / Notes |
|---|---|---|---|---|---|---|---|---|
| 1 | Gate A | 1 | ShortestPathMapBasedMovement | 0 | 0 | 5M | Bluetooth | Cluster at (1200,1000) |
| 2 | Gate B | 1 | ShortestPathMapBasedMovement | 0 | 0 | 5M | Bluetooth | Cluster at (1300,1050) |
| 3 | Gate C | 1 | ShortestPathMapBasedMovement | 0 | 0 | 5M | Bluetooth | Cluster at (1400,1100) |
| 4 | Shops | 25 | ShortestPathMapBasedMovement | 0 | 0 | 5M | Bluetooth | Cluster at (1200,1000) |
| 5 | Restrooms | 20 | ShortestPathMapBasedMovement | 0 | 0 | 3M | Bluetooth | Cluster at (1300,1050) |
| 6 | Pedestrians A | 67 | ShortestPathMapBasedMovement | 1-3 | 0-120 | 5M | Bluetooth | Cluster at Gate A, random start |
| 7 | Pedestrians B | 67 | ShortestPathMapBasedMovement | 1-3 | 0-120 | 5M | Bluetooth | Cluster at Gate B, random start |
| 8 | Pedestrians C | 66 | ShortestPathMapBasedMovement | 1-3 | 0-120 | 5M | Bluetooth | Cluster at Gate C, random start |
| 9 | Security | 15 | MapBasedMovement | 1-3 | 10-60 | 25M | Bluetooth | Random start, patrolling areas |
| 10 | Staff | 40 | ShortestPathMapBasedMovement | 1-3 | 10-60 | 15M | Bluetooth | Random start, general staff movement |

Fig. 8. Node groups

*4) Performance metrics*

To assess the performance of the routing protocols, we will compare Epidemic routing and SNF based on key performance metrics:

**Delivery Ratio:**

Represents the fraction of messages successfully delivered to the destination out of all messages generated.

**Relevance:**

In this scenario, a high delivery ratio indicates that the network reliably delivers these critical "HELP" messages from the victims to the security, which could be lifesaving in real-world situations.

**Latency:**

Measures the average time taken for a message to reach its destination after being generated.

**Relevance:**

Timely delivery is also important for this scenario. Low end-to-end delay ensures that the "HELP" messages reach security quickly.

## VI. SIMULATION RESULTS

### A. Simulation 1: Base comparison of Epidemic and SNF

In this simulation we compare Epidemic and SNF with the configuration from figure 8. The scenario follows a setup, where security walks random between pedestrians, who are moving towards gates.

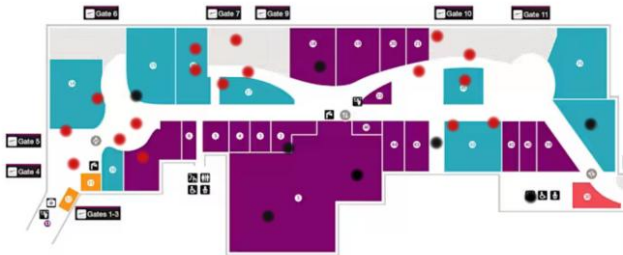

Fig. 9. Security (black dots), pedestrians (red dots)

The results show that Spray-and-Focus (SnF) performs significantly better than Epidemic. SnF achieves a significantly higher delivery ratio (0.99) compared to Epidemic (0.75). Additionally, SnF exhibits a drastically lower latency (0.19 s), meaning that help messages reach their destinations far faster than with Epidemic (3597 s ≈ 60 min).

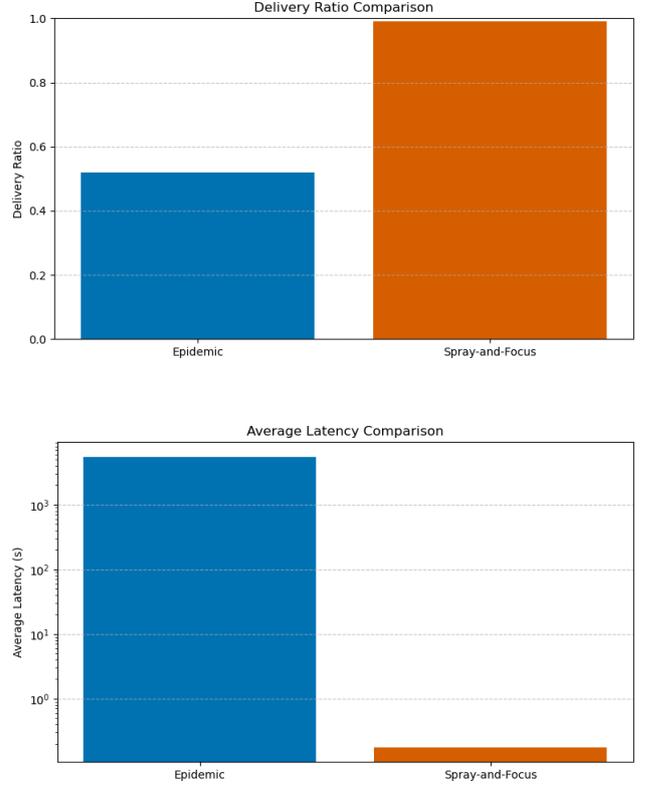

Fig. 10. Bar chart showing delivery ratio and average latency

### B. Simulation 2: Increasing security nodes

In this simulation, we try to increase the security nodes to see how it affects the performance of the protocols; can we increase delivery of messages from victims and lower latency by increasing more security?

In the Epidemic scenario, there is generally a linear tendency where increasing the number of security nodes improves delivery ratio and reduces latency (3207 s ≈ 53 min) at batch 5 (security nodes = 14). However, batch b3 (12 security nodes) appears as an outlier at batch b4.

In SnF, increasing the number of security nodes has less effect on the delivery ratio, which remains near-perfect (0.99). Latency shows minor fluctuations and does not consistently decrease with more security nodes; in fact, an increase is seen averaging 1 s.

This suggests that once coverage is sufficient, adding extra nodes provides minimal improvement for SnF, in contrast to Epidemic routing where additional nodes noticeably impact both delivery and latency.

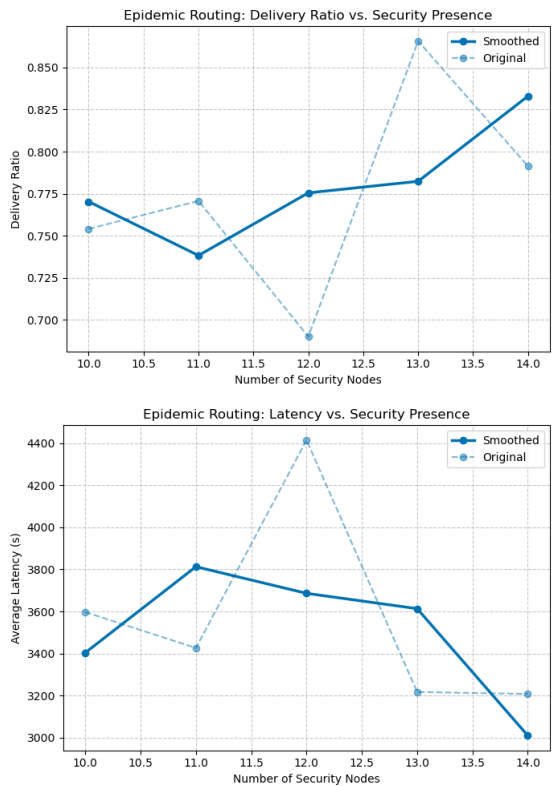

Fig. 11. Epidemic: Line chart of increasing security nodes

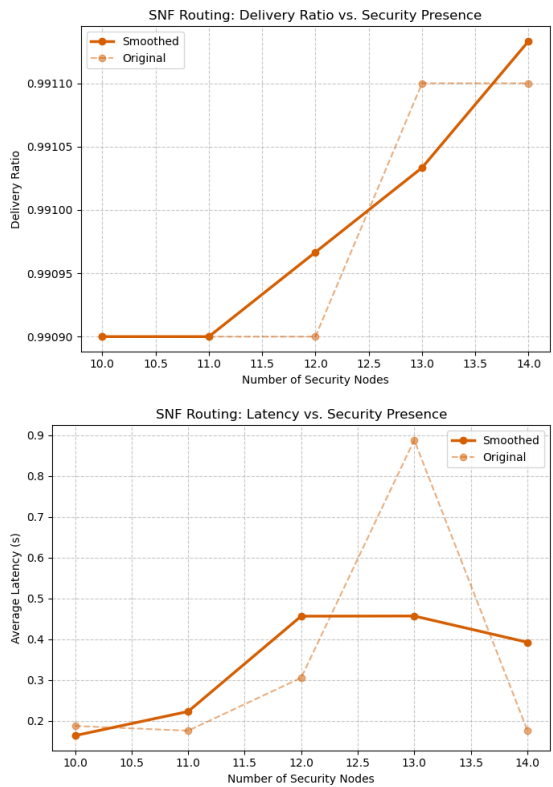

Fig. 12. SNF: Line chart of increasing security nodes

## C. Simulation 3: Modifying movement patterns

To investigate whether targeted mobility can improve message delivery, we modify the movement of the security to target the gates. The scenario follows a setup, where security walk random between nodes.

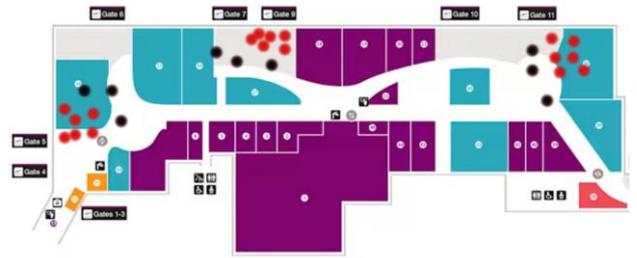

Fig. 13. Security (black dots), pedestrians (red dots)

In the modified mobility configuration, only security nodes were adjusted. Originally, security moved more randomly along the simulation map using MapBasedMovement. In the new setup, security is split into three groups, each assigned a cluster center near one of the main gates. This ensures that their movement is concentrated around the gates where pedestrians naturally converge.

Each security group patrols or clusters around its designated gate, maintaining a defined zone of activity, which aligns their presence with pedestrian flow and improves the likelihood of timely alert message delivery.

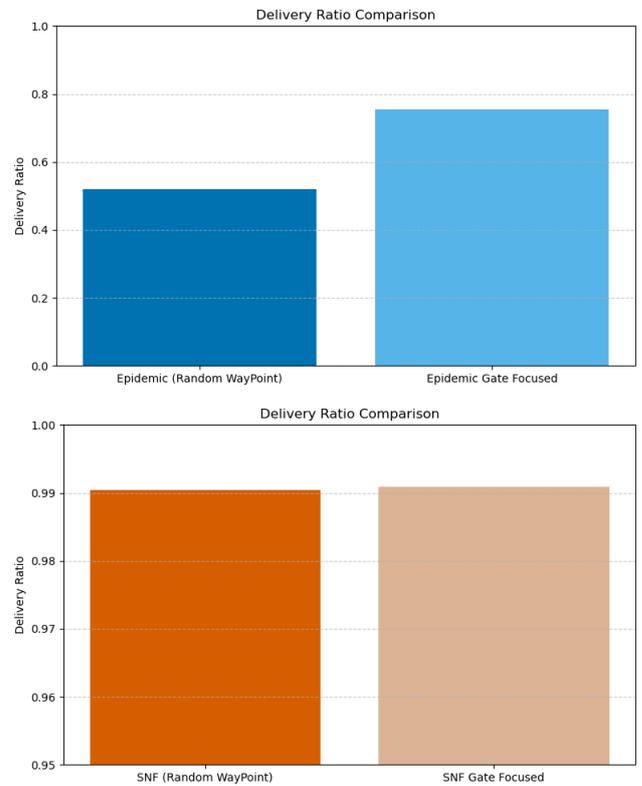

In the Epidemic scenario, positioning security nodes at pedestrian hotspots significantly improved performance compared to random map-based patrol. Specifically, the delivery ratio increased from 0.52 s with random map-based movement to 0.75 s when security nodes focused on gates, approaching the highest batch performance of 0.79. Average latency similarly decreased from 5523 s ($\approx$ 92 min) to 3597 s ($\approx$ 60 min), matching the latency of the highest batch.

For Spray-and-Focus, directing security nodes toward gates had strictly limited performance. Delivery ratios were consistently high across configurations: 0.99 for map-based movement, 0.99 for gate-focused, and 0.99 for the highest batch. Latency remained significantly low, between 0.18 s and 0.19 s.

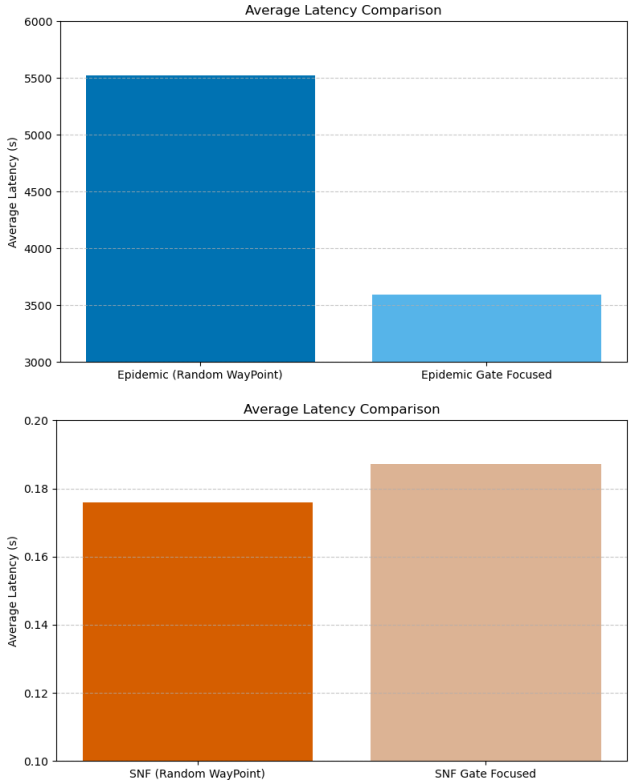

Fig. 14. Bar charts showing random movement and focused movement towards gate for Epidemic and SnF

These results indicate that, for Epidemic routing, strategic positioning of security nodes near pedestrian clusters can achieve almost the same benefit as increasing the number of nodes. For Spray-and-Focus, performance is already near optimal, so ensuring mobility across security nodes is sufficient to maintain peak delivery and minimal latency.

## VII. CRITICAL DISCUSSION OF SIMULATION RESULTS

### A. Performance of Epidemic and SNF

The results show that the success of delivering alert messages from victims to security personnel depends more on the placement and movement of security nodes than on simply increasing their number.

For the Epidemic protocol, directing security nodes toward pedestrian clusters achieved performance comparable to scenarios with the maximum number of security nodes. Average latency decreased from 5523 s ($\approx$ 92 min) to 3597 s ($\approx$ 60 min), and delivery ratios increased from 0.52 to 0.75, demonstrating the strong impact of strategic positioning. This occurs because Epidemic relies on opportunistic forwarding; placing nodes where pedestrian traffic is dense increases the likelihood of encounters and message propagation.

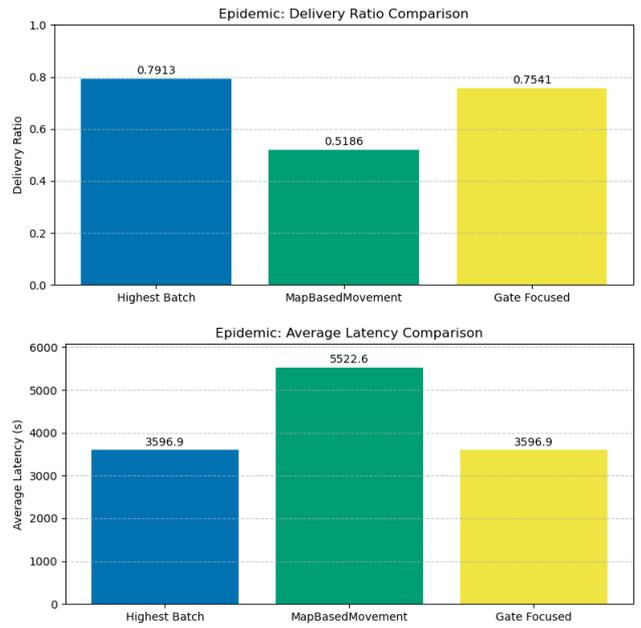

Fig. 15. Epidemic: Bar chart showing comparison of all 3 simulations

For Spray-and-Focus (SnF), performance remained consistently high regardless of the number of security nodes, with delivery ratios around 0.99 s and low latency ( 0.18 s). SnF uses a controlled number of message copies combined with targeted forwarding toward recently encountered nodes, which makes it achieve high performance even without changes in node positioning.

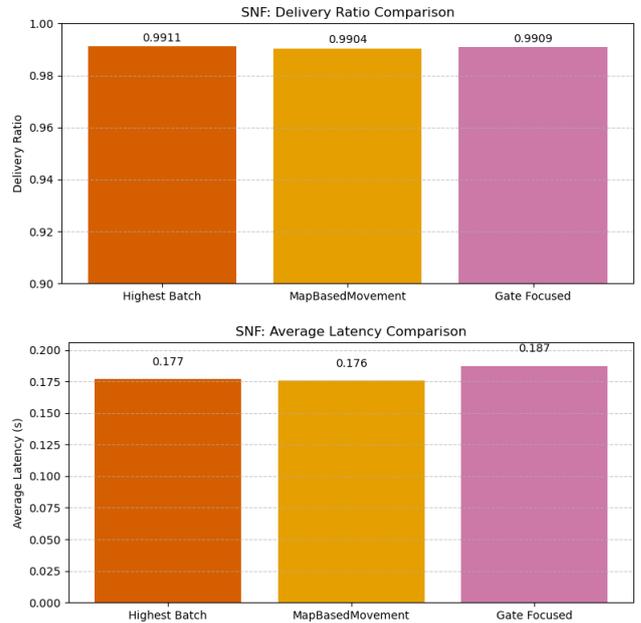

Fig. 16. SNF: Bar chart showing comparison of all 3 simulations

The key takeaway is that adjusting mobility patterns to focus on areas where pedestrians naturally converge can significantly improve performance for Epidemic routing. Strategically redistributing existing resources can achieve similar or better results than adding additional personnel when utilizing either SnF or Epidemic.

## B. Optimizing mobility and security to combat human trafficking via DTNs

In real-world airport environments, passenger movement is inherently dynamic and unpredictable, making rigid control over how they move impractical. However, there are various technologies used in airports that gather data about passengers, as shown in the study by Alabsi and Gill (2021) [1]. IoT devices and sensors are placed throughout key airport areas, such as gates, walkways, and shops, to monitor passenger movement[1]. RFID technology and mobile applications enable real-time tracking of passenger locations and luggage, helping airports manage baggage handling efficiently and provide guidance to passengers during their travel journey[1]. Biometric systems are used at critical stages such as check-in, security control, border control, and boarding, capturing identity information to streamline processing and improve passenger experience while also generating valuable operational data [1].The study also highlights the ethical and privacy considerations associated with collecting and using these data. Challenges include unauthorized access, secondary use of information, and potential leakage[1]..By leveraging all the latter described data, we may anticipate high-traffic areas and strategically position security nodes to maximize contact with pedestrians. By leveraging the types of data described above, airports may anticipate high-traffic areas and strategically position security personnel to maximize contact with travelers. For airport authorities and policymakers, this suggests that optimizing the placement and mobility of current resources can enhance safety and operational efficiency while potentially reducing costs.

## C. Limitations of simulation and simulating in ONE

A key limitation of using ONE for simulation is the lack of real-world data traces, particularly regarding movement within an airport. While some node groups are known, we do not have complete coverage of all possible groups, nor do we have data on their relative proportions. For example, we know from airport layouts that there are gates, shops as illustrated in figure 5. However, airports also contain other mobile entities, such as trams, service vehicles, luggage carts, and staff with specialized roles, which we cannot fully account for. Other node groups, therefore, remain unknown. Despite these gaps and uncertainties, the simulation can still provide meaningful insights. By modeling behaviors for the known nodes, ONE allows us to explore the general dynamics of how delay tolerant protocols might operate, identify potential bottlenecks, and evaluate relative performance under different scenarios[11].

## VIII. DISCUSSION OF DTNS AND RELATED TECHNOLOGIES

### A. Existing technologies to combat human trafficking

Prior work has developed machine learning pipelines to detect potential human trafficking in online escort advertisements[9]. Ads are analyzed to assign risk scores, enabling investigators to prioritize case studies efficiently. This approach has been integrated into the Memex platform, which indexes over 100 million ads and is actively used by more than 200 law enforcement agencies.

Another study also explored the use of computer vision to aid child trafficking investigations by identifying school uniforms in images[13]. A two-stage pipeline classifies whether a uniform is present and predicts attributes such as color and type to infer the child's school of origin. A more in situ intervention is demonstrated by Love Justice International demonstrates a proactive, technology-assisted approach to combat human trafficking by intervening before victims are exploited[12]. The organization combines historical victim data, OpenStreetMap road networks, and socioeconomic data from NASA to generate predictive route heatmaps identifying transit routes. Transit monitoring stations are placed along these routes, where trained staff interview travellers and use machine learning models to determine risks. High risk individuals are either guided to safety or referred to law enforcement, resulting in over 30,000 interceptions and more than 1,100 arrests by this initiative.

### B. Tools for at-risk individuals

As highlighted in the motivation, human trafficking affects millions of people worldwide and is likely to intensify as conflicts, crises, and displacement continue to escalate. Despite the demonstrated effectiveness of current technologies, further potential exists to deliver real-time support directly to at-risk individuals, particularly within high-risk transit hubs such as airports. Victim-facing tools deployed on personal devices could enable timely alerts and requests for assistance. Such systems could be pre-installed on smartphones, similar to Find My iPhone location sharing[2], or Safety Check In[3] Safety check in enables the user to set a time slot on arriving at a specific location, where the app automatically notifies a contact to a user, that it has arrived at that location, at the specified time. If the contact doesn't receive a notification within a certain amount of time after the defined arrival time, something may be wrong. These systems ensure immediate availability without requiring users to download new applications.

### C. Delay-Tolerant Networks and policy considerations

Building on the concept of these victim-facing tools on personal devices, delay-tolerant networks (DTNs) may provide a practical solution in situations where real-time connectivity is limited. By routing alerts locally through nearby devices, DTNs may allow messages from at-risk individuals to reach authorities even when cellular or Wi-Fi access is unavailable, providing the means for immediate action in high-risk environments like airports. As outlined in 5.1,, governments and international organizations, including the UNODC, IATA, and the EU are highlt encouraging industry-level anti-trafficking measures. While DTN-based alert systems are technically feasible, the main challenges may lie in policy implementation and operational deployment rather than the technology itself.

## IX. CONCLUSION

With an estimated 25 million people trafficked globally in 2018, and approximately 80% of them passing through airports, these transit hubs are critical points for intervention. At-risk individuals in airports face challenges in communicating reliably due to cellular coverage gaps, paid or restricted Wi-Fi, and network congestion. This work demonstrates how Delay-Tolerant Networks (DTNs) can help ensure timely and reliable delivery of alerts from potential victims to security.

Simulation results indicate that simply adding more security nodes provides limited benefits. Greater improvements are achieved by high mobility. For both Epidemic, aligning node placement with clusters, e.g., when pedestrians cluster towards a gate, significantly improved message delivery ratio from 52% to 75%. In particular, SnF maintained consistently high delivery ratios around 99% while keeping latency significantly low ( 0.18 s). These findings suggest that by applying DTNs, optimizing the placement of existing security resources can achieve similar or better performance than increasing security. First, it helps airport operators maintain strong coverage without increasing costs. Most importantly, the results highly indicate that victims' alerts reach authorities quickly and reliably.

Including real passenger flow data, both in the simulator and when developing a DTN system for this scenario, could improve the results and make the simulations more realistic and useful in practice.

Delay-Tolerant Networks (DTNs) provide a valuable complement to existing anti-trafficking measures by enabling timely, local alerts and ensuring critical messages reach authorities.